\documentclass[%
 reprint,
 amsmath,
 amssymb,
 aps,
]{revtex4-1}

\usepackage{graphicx}
\usepackage{subfigure}
\usepackage{mathrsfs}
\usepackage{caption}
\usepackage{latexsym}
\usepackage{bm}
\usepackage{float}

\begin{document}
\preprint{APS/123-QED}
\title{Quantum Annealing for Semi-Supervised Learning}
\author{Yu-Lin Zheng}
\author{Wen Zhang}
\author{Cheng Zhou}
\author{Wei Geng} \email{wei.geng@huawei.com}

\affiliation{
 Hisilicon research, Huawei Technologies Co., Ltd., Shenzhen, China
}

\date{\today}

\begin{abstract}
Recent advances in quantum technology have led to the development and the manufacturing of programmable quantum annealers that promise to solve certain combinatorial optimization problems faster than their classical counterparts.
Semi-supervised learning is a machine learning technique that makes use of both labeled and unlabeled data for training, which enables a good classifier with only a small amount of labeled data. 
In this paper, we propose and theoretically analyze a graph-based semi-supervised learning method with the aid of the quantum annealing technique, which efficiently utilize the quantum resources while maintaining a good accuracy.
We illustrate two classification examples, suggesting the feasibility of this method even with a small portion (20\%) of labeled data is involved.
\end{abstract}
\maketitle

\section{Introduction}
The recent developments of machine learning, enable computers to infer patterns that were previously untenable from a large data set ~\cite{michie1994machine, bishop2006pattern}. Quantum computing, on the other hand, has been proved to outperform classical computers in some specific applications ~\cite{steane1998quantum, hirvensalo2013quantum}. To extend both advantages, increasing interests have been made to explore the merging of these two disciplines \cite{schuld2015introduction,adcock2015advances,biamonte2017quantum}. For instance, the quantum version of linear models of machine learning, such as support vector machines (SVM) ~\cite{rebentrost2014quantum}, principal component analysis (PCA) ~\cite{lloyd2014quantum}, can be potentially more efficient than their classical versions. Quantum generative models were also proposed with exponential speedups than the traditional models ~\cite{gao_quantum_2018}. However, most of those algorithms require a large-scale fault-tolerant quantum computer that is beyond the ability of current hardware techniques.
Meanwhile, quantum annealer, as one of the noisy-intermediate scale quantum (NISQ) devices ~\cite{preskill2018quantum}, has been proved useful in many applications such as optimization ~\cite{neukart2017traffic}, simulation ~\cite{babbush2014adiabatic} and machine learning ~\cite{adachi2015application}. 
In this work, we propose a method to tackle semi-supervised classification tasks on a quantum annealer.
An encoding scheme and a similarity-calculation method that map the graph representation of the problem to the Hamiltonian of a quantum annealing (QA) system are suggested, which avoid the implementation of multi-qubit interaction. We show in two examples that good classification accuracies can be achieved using only a small amount of labeled data.

\subsection{Semi-Supervised Learning}

Semi-supervised learning (SSL) method usually refers to classification tasks where trainings are usually carried out with both labeled and unlabeled data when only a small amount of labeled data is available \cite{chapelle2009semi,zhu2005semi, zhou2014semi}. A well-known model of this kind is the graph-based method, by which all the data are represented by vertices of a graph, and similarities between data are represented by the edges of the graph ~\cite{zha2009graph}.
Specifically, we are given a labeled data set $\mathcal{D}_{l} =\{(\mathbf{d}_1, lab_1), (\mathbf{d}_2, lab_2), \dots, (\mathbf{d}_l, lab_l)\}$ and an unlabeled data set $\mathcal{D}_{u} = \{\mathbf{d}_{l+1}, \mathbf{d}_{l+2}, \dots, \mathbf{d}_{l+u}\}$, where $lab_i$ is the label of data $\mathbf{d}_i$, $l$ and $u$ are the numbers of labeled and unlabeled data, respectively.
In most cases, the data set has fewer degrees of freedom than its original dimension, which allows us to map the data $\mathbf{d}_i$ from its original space to $\mathbf{x}_i$ in a certain manifold after some dimension-reduction processes, such as locally linear embedding ~\cite{roweis2000nonlinear} and principal component analysis ~\cite{kambhatla1997dimension}. 
Therefore, a graph $G=(V, S)$ can be built based on $\mathcal{D} = \mathcal{D}_l\cup \mathcal{D}_u$, in which the set of vertices $V = \{\mathbf{x}_1, \dots, \mathbf{x}_{l+1}, \dots, \mathbf{x}_{l+u}\}$ represents all the data vectors in the manifold. The adjacency matrix $S = [S_{ij}]$ of this graph represents the similarities between the $i$-th data and the $j$-th data for $1\leq i,j \leq l+u$. Here we assume a symmetric similarity, i.e., $S_{ij}=S_{ji}$ for $i \neq j$ and $S_{ii}=0$.
During the training process, label information tends to propagate along edges with greater similarities. When the training process ends, labels of the unlabelled vertices are decided according to their probability distributions.

In practice, we need to deal with integer programming and combinatorial optimization when solving most SSL problems, which are usually non-convex or non-smooth ~\cite{zhou2014semi}. In order to obtain a global optimal solution, the solving process usually involves high time and space complexities. With the potential speed-up in solving the optimization problems, the quantum annealing method is a natural consideration to be applied in SSL.

\subsection{Quantum Annealing}
In a QA process~\cite{kadowaki1998quantum}, the system is firstly prepared in a ground state of an initial Hamiltonian. A target Hamiltonian is gradually applied to the system as it evolves following the time-dependent Schr\"odinger equation. 
If the application of the target Hamiltonian is slow enough, the system will adiabatically stay at the ground state of the instantaneous Hamiltonian and finally reach to the ground state of the target Hamiltonian, which encodes the solution of the problem.
Demonstration of QA have been vastly reported using systems based on superconducting circuits ~\cite{DWave1989,bunyk2014architectural,boixo2013experimental,boixo2014evidence}. 

When an Ising model is used in QA system, the Hamiltonian of the annealing process is usually defined as below:
\begin{equation}
    H=s(t)H_{\mathrm{ini}}+\left(1-s(t)\right)H_{\mathrm{tar}}
\end{equation}
\begin{equation}
\label{HQA}
H_{\mathrm{tar}}=-\sum_{i=1}^{N}h_i \sigma_i-\sum_{i,j=1, i\neq j }^{N}J_{ij}\sigma_i \sigma_j,
\end{equation}
in which $H_{\mathrm{ini}}$ and $H_{\mathrm{tar}}$ are the initial and target Hamiltonian of the system, respectively, $h_i$ is the bias applied on the $i$-th qubit, $\sigma_i$ is a Pauli Z operator on the $i$-th qubit and $J_{ij}$ is the coupling parameter between the $i$-th qubit and the $j$-th qubit. 
$s(t)$, as a function of time $t$, controls the annealing speed by monotonically decreasing from $1$ to $0$.

\section{Method}
\label{sct&agr} 

\subsection{Label Encoding}
\label{lb_ecd}
The graph $G$ can be mapped onto a QA system by associating each vertex $i$ with a group of $\alpha$ qubits $\{q_i^{(1)},q_i^{(2)},\dots,q_i^{(\alpha)}\}$ that encodes the label, and by associating the edge $S_{ij}$ between vertices $i$ and $j$ with the coupling strengths $J_{ij}$ between qubit groups. The binary nature of qubits leads to an intuitive choice of a binary encoding for labels, which allows the group to encode up to $K=2^\alpha$ labels. Such an encoding scheme usually calls for global constraints on the group of qubit in order to guarantee that a proper code is simultaneously attributed to the whole group at the end of the annealing process. However, this can only be achieved via multi-qubit interactions, a mechanism that current QA hardware can hardly address. One-hot encoding can overcome this requirement by reducing the interaction to a quadratic order at a cost of exponentially increased qubit resources with regard to the label number compared with the binary encoding \cite{kumar2018quantum, ulanov2019quantum}.

Here we propose another encoding method that essentially uses binary codes while avoiding the the hardware complexity introduced by multi-qubit interactions. We suggest that the $a$-th bit in the binary label code of all unlabeled data is determined independently by a QA process, which consists of the qubits $\{q_1^{(a)},q_2^{(a)},\dots,q_{l+u}^{(a)}\}$. In this case, the global information shared within one group of qubit should be maintained by the consistency of similarity between arbitrary two data and the Hamming distance of their labels. Specifically, we first calculate the barycenters of each label on the manifold by $\mathbf{X}^{(k)}=\frac{1}{n_k}\sum_i\mathbf{x}_i^{(k)}$ according to the labeled data set $\mathcal{D}_{l}$, where $n_k$ denotes the number of data with the $k$-th label. A shortest path that visits all the barycenters is then searched in the manifold, leading to a sequence of all the barycenters: $\{\mathbf{X}^{(k)}|k\in\mathcal{K}\}$, where $\mathcal{K}$ is the set of all labels. According to this sequence, labels are assigned with an ordered Gray code, which ensures that only one bit changes in the codes of two adjacent labels. Thus, the correlation of distances between arbitrary data and their label codes' Hamming distance is optimized.
Though the complexity of finding the shortest path in the manifold is equivalent to the well-known travelling salesman problem, in most cases, the number of label is far fewer than that of data in a given data set. It has also been shown that this kind of task could also be potentially accelerated by a QA device ~\cite{martovnak2004quantum}. 

There are certainly cases that $2^{\alpha-1}<K<2^\alpha$. To avoid that the redundant codes are wrongly attributed to an unlabeled data, we can use up to two codes to encode one label while assuring the two codes are next to each other in the ordered Gray code. 
For example, if a group of $\alpha = 3$ qubit is used to encode only $K = 5$ labels, the label codes can be attributed as follows: $\{000, 001\}_{\mathrm{label}\:1}$, $\{011, 010\}_{\mathrm{label}\:2}$, $\{110, 111\}_{\mathrm{label}\:3}$, $\{101\}_{\mathrm{label}\:4}$, $\{100\}_{\mathrm{label}\:5}$.

\subsection{Structure of the system}
Based on the aforementioned encoding method, the whole training of the SSL classification task can be divided into $\alpha$ independent layers, of which the $a$-th corresponds to an annealing process of qubits $\{q_1^{(a)},q_2^{(a)},\dots,q_{l+u}^{(a)}\}$. Because of the limited connectivity that a QA hardware can currently achieve, we only require that each qubit is logically coupled with at least $\xi$ qubit in each layer. $\xi$ may depends on the certain distribution of a data set and should be a small number compared with the total amount of data.

\begin{figure}[t]
\centering
\includegraphics[width = .4\textwidth]{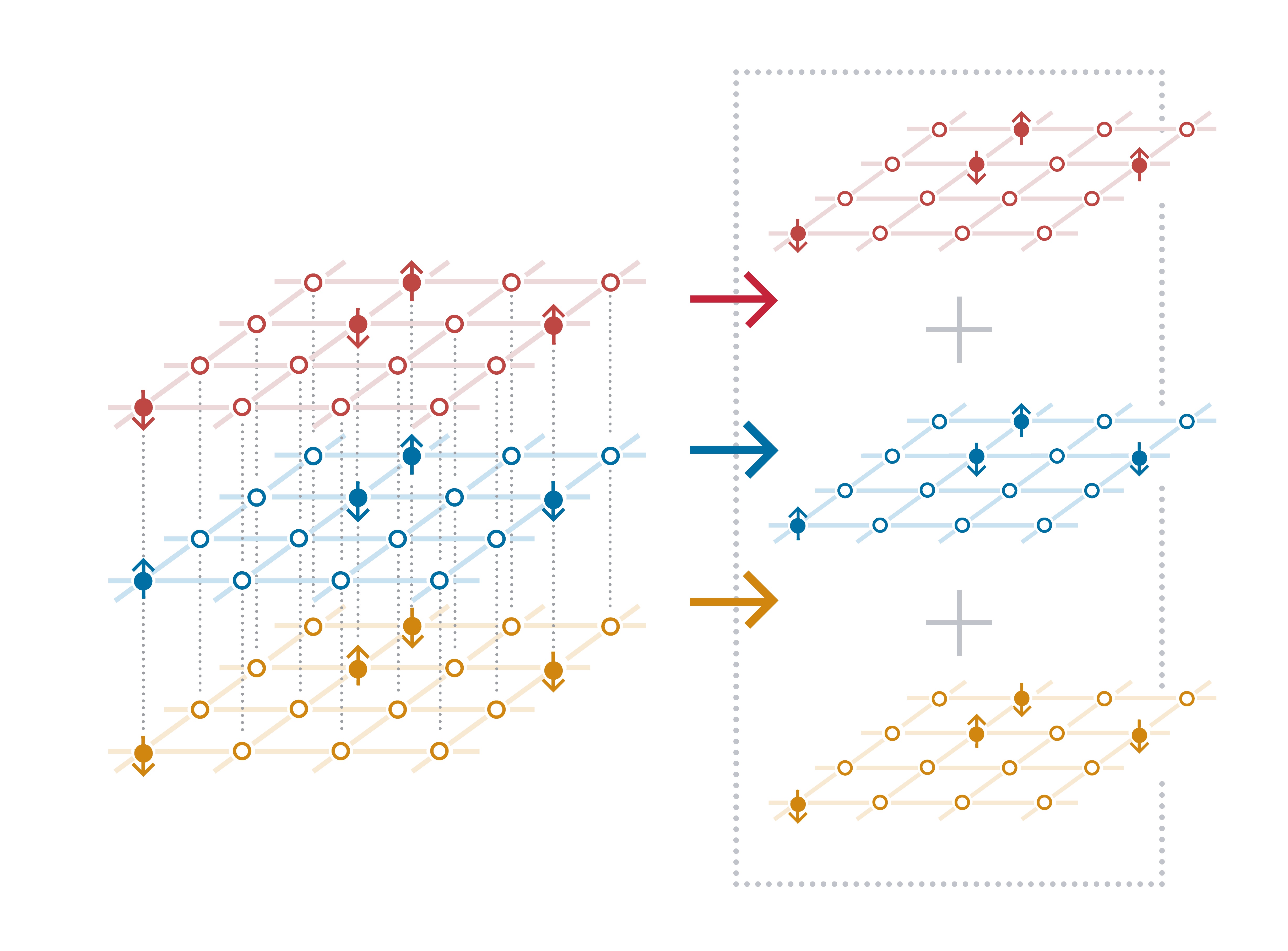}
\caption{\label{fig:structure}
Example of the QA structure that performs the SSL classification task. Here $\alpha = 3$. Each qubit, depicted in the solid or open circle, expresses one bit of the label code of a labeled or unlabeled data, respectively. A group of three qubits connected with a dashed line represents one data. Arrows on the labeled data indicate the directions of $h_i$ on corresponding qubit $q_i$. In this example, each qubit in the same layer is topologically coupled with its 4 neighbors. A time-division multiplexing scheme can be used by dividing the system in to 3 smaller systems that are annealed individually.}
\end{figure}

This system can naturally lead to a time-division multiplexing manner, such that each part of the training process can be operated separately in time using just one smaller system. This is especially advantageous when the number of qubit in a QA hardware is limited compared with the problem size. In fact, such a time-division multiplexing manner is equivalent to a dichotomy method, which is, by determining each bit of the binary label code, the total unlabeled data are sorted into two groups after each annealing process. An example of such a system is delineated in Fig.~\ref{fig:structure}. 

Moreover, we specify two configurations for labeled and unlabeled data separately:
\textbf{Labeled data}: To assure that the qubits of labeled data reveal correct labels after being measured, we should apply a bias $h_i$ that is large enough to make the probability of their transition to wrong states close to $0$ at the end of the QA process.
\textbf{Unlabeled data}: No bias is applied to the corresponding qubits.

Hence, Eq. \ref{HQA} can be re-written as:
\begin{equation}
\label{SSL-HQA}
    H_{\mathrm{tar}}=-\sum_{a=1}^{\alpha}\left(\sum_{i=1}^{l}h_i^{(a)} \sigma_i^{(a)} + \sum_{i,j=1}^{l+u}J_{ij}\sigma_i^{(a)} \sigma_j^{(a)}\right).
\end{equation}
Here $\sigma_i^{(a)}$ is the Pauli Z operator on the $i$-th qubit at the $a$-th layer in the system. 
After the annealing process, we can obtain the labels by measuring the corresponding qubits with its Pauli Z operator. 
The label we obtained is written as $y_i=\mathrm{string}\:y_i^{(1)}\dots y_i^{(\alpha)}$. In order to describe labels in quantum language, we use $y_i^{(a)} \in \{-1, +1\}$ rather than $\{0, 1\}$ which depends on the state after being measured denoted as $|y_i^{(a)}\rangle$, in which $|y_i^{(a)}\rangle \langle y_i^{(a)}| = \frac{1}{2}\left(1+y_i^{(a)}\sigma_i^{(a)}\right)$.

\subsection{Similarity and coupling parameters}
In the QA model of Eq. \ref{SSL-HQA}, when $J_{ij}>0$, the stronger the two qubits are coupled, the more likely they are to have the same orientations. Therefore, it is intuitive to map the similarity between two data to the coupling coefficient between two qubits in a QA system.
According to the vectors of two data in the manifold, the similarity between the two data can be calculated as below:
\begin{flalign}
\label{similarity_matrix} 
S_{ij}=\left\{\begin{matrix}
f(\|\mathbf{x}_i-\mathbf{x}_j \|_{p}) &,& \mathrm{if} \quad i\neq j\\
0 &,& \mathrm{otherwise}
\end{matrix}\right.,
\end{flalign}
where $\|\mathbf{\Theta}\|_p$ is the p-norm of vector $\mathbf {\Theta}$ and $f(\Theta)$ is a monotonically decreasing function of $\Theta$. To better describe the similarities of a particular data set, $f(\Theta)$ may contain parameters that can be learned as will be discussed in the next section.
For example, we can use Euclidean distance-based similarity:
\begin{flalign}\label{Similarity-Eucli_Dis}
S_{ij} = \left\{\begin{matrix}
\frac{\beta_1}{\beta_2+\|\mathbf{x}_i-\mathbf{x}_j\|_2} &,& \mathrm{if} \quad i\neq j\\
0 &,& \mathrm{otherwise}
\end{matrix}\right.,
\end{flalign}
where $\beta_1$ and $\beta_2$ are parameters to be learned. 

For some data that have normal distributions in the manifold, we can also use the Gaussian-like function to describe the similarity between $\mathbf{x}_i$ and $\mathbf{x}_j$ as: 

\begin{flalign}
\label{weight_gaussian} 
S_{ij}=\left\{\begin{matrix}
\sum\limits_{k=1}^{K} \frac{p(k)}{\sqrt{(2\pi)^d|B_k|}}e^{-\frac{1}{2}(\mathbf{x}_i-\mathbf{x}_j)^T B_k^{-1}(\mathbf{x}_i-\mathbf{x}_j)} &, & \mathrm{if} \quad i\neq j\\
0 &, & \mathrm{otherwise}
\end{matrix}\right.,
\end{flalign}
in which $B_k$ is assumed to be a symmetric matrix whose elements are to be learned from the labeled data set $\mathcal{D}_l$, $d$ is the dimension of the manifold, $p(k)$ is the proportion of the $k$-th label in a given labeled data set, i.e., $p(k) = \frac{n_k}{l}$ in which $n_k$ is the amount of data with the $k$-th labels.

We then map the similarity to the coupling parameter in a QA system. As mentioned above, a limited connectivity of a qubit with others in the system is usually favoured to reduce the hardware complexity as well as to increase the accuracy. The latter is due to the consideration that an unlabeled data should be affected mainly by the surrounding ones in the manifold which are more likely to have smaller Hamming distance.
The configuration of connection can be determined using a symmetric mask matrix $M$ applied on the similarity matrix $S$. First, we search the $\xi$ largest elements in row $S_{i*}$ to find the $\xi$ closest ones to data ${\bf x}_i$. A matrix $M'$ is defined as:

\begin{flalign}
M'_{ij}=\left\{\begin{matrix}
1 &, & \mathrm{if} \: S_{ij} \: \textrm{is among the} \: \xi \: \textrm{largest of} \: S_{i*}
\\
0 &, & \mathrm{otherwise}
\end{matrix}\right..
\end{flalign}
The mask matrix is then calculated by $M = (M')^T \mathrm{OR} (M')$, in which $(A) \mathrm{OR} (B)$ means take bitwise OR operation between matrix $A$ and $B$.
The coupling parameters $J_{ij}$ between two arbitrary qubits in Eq. \ref{SSL-HQA} can be calculated as:

\begin{equation}
\label{coupling matrix}
    J_{ij} = M_{ij}*S_{ij}.
\end{equation}

It should be noted that there may be cases where more than $\xi$ qubits need to be topologically coupled to a certain one using this approach. A random elimination of the coupling parameters can be made to further reduce the connectivity. Or we can use extra qubits as intermediate couplers to realize a larger connectivity at a cost of reduced problem size that can be solved ~\cite{adachi2015application}.

\subsection{Parameters learning}

The final step concerns the attribution of appropriate values to the parameters that are related to the system's Hamiltonian. Firstly, parameters involved in the similarity calculation can be determined by a supervised learning process using the labeled data set. 
In the learning process, $H_{\mathrm{tar}}$ is replaced with $H_{\mathrm{learn}}$ by removing the bias $h_i$, thus we have
\begin{equation}
    H_{\mathrm{learn}}= -\sum_{a=1}^{\alpha}\ \sum_{i,j=1}^{l}J_{ij}\sigma_i^{(a)} \sigma_j^{(a)}.
\end{equation}
$E_{\mathrm{learn}}$ is the system energy corresponds to $H_{\mathrm{learn}}$. 
The probability distribution of labels after the QA process follows the Bolzman distribution as:
\begin{equation}
\label{P-function}
\begin{aligned}
    &p(y_1,\dots,y_l|\mathbf{x}_1,\dots,\mathbf{x}_l;\mathbf{\theta})\\
    =&\frac{1}{Z}\exp\left(-\frac{E_{\mathrm{learn}}(y_1,\dots,y_l)}{T}\right)\\
    =&\frac{1}{Z}\exp\left(-\frac{\sum_a^{\alpha}\sum_{i,j=1}^{l}\langle y_i^{(a)}y_j^{(a)}|H_{\mathrm{learn}}|y_i^{(a)}y_j^{(a)} \rangle}{T}\right)\\
    =&\frac{1}{Z}\exp\left(-\frac{\sum_a^{\alpha}\sum_{i,j=1}^{l} -J_{ij}y_i^{(a)}y_j^{(a)}}{T}\right)
\end{aligned}
\end{equation}
where $Z$ is the partition function and $Z=\mathrm{Tr}[\exp\{-H_{\mathrm{learn}}/T\}]$, $\mathbf{\theta}$ is the vector containing all parameters to learn. 

A negative log-likelihood function is therefore defined as below:
\begin{equation}\label{L-function}
 \mathcal{L}(\mathcal{D}_l;\mathbf{\theta}) 
           = -\log p(lab_1,\dots,lab_l|{\mathbf x}_1,\dots,\mathbf{x}_l;\mathbf{\theta}).
\end{equation} 

Then the elements of $\mathbf{\theta}$ are determined by minimizing the conditional likelihood separately:
\begin{equation}
\begin{aligned}\label{P-learning}
    \theta_j = \arg \min -\log p(lab_1,\dots,lab_l|{\mathbf x}_1,\dots,\mathbf{x}_l;\mathbf{\theta}).
\end{aligned}
\end{equation}

Such a learning process is similar with the Boltzmann machine model ~\cite{hinton1986learning,amin2018quantum,ulanov2019quantum}, except that the sampling process can be accelerated by iterated QA processes and project measurements of qubits ~\cite{adachi2015application}.

Again, some hyperparameters needs to be tuned properly according to a specific data set and hardware system. For instance, the value of $h_i^{(a)}$ should be large enough such that the labels of labeled data is obtained correctly after annealing process. $\xi$ should be properly set to (i) ensure that every unlabeled data can reach to a labeled one through a connected path to avoid a random result and to (ii) reduce the implication of data that are far away in the manifold.

\section{Example}
Here we give two examples based on realistic database to verify the method discussed above. As a proof-of-principle demonstration, the annealing processes are simulated by a classical computer.

\begin{figure}
\centering
    \includegraphics[width = 0.4\textwidth]{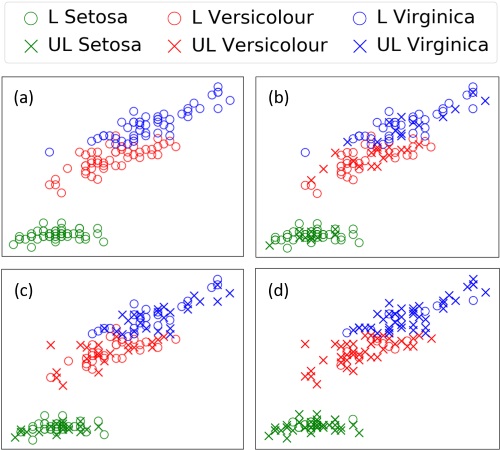}
    \caption{The iris data set (a) and the classification results using the algorithm proposed in this work when the portion of unlabeled is (b) 30\%, (c) 50\% and (d) 80\%. }
    \label{fig:iris}
\end{figure}

\begin{figure}
\centering
    \includegraphics[width = 0.4\textwidth]{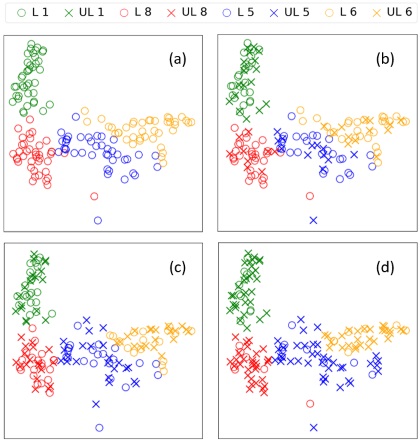}
    \caption{The handwriting digital data set with reduced dimensions (a) and the classification results using the algorithm proposed in this work when the portion of unlabeled is (b) 30\%, (c) 50\% and (d) 80\%.}
\label{fig:handwriting digits}
\end{figure}

\subsection{Example 1: Iris}
We first use a database of iris that has been widely used in pattern recognition literatures~\cite{fisher1936use}.
There are three kinds of label in the data set, shown by points in three colors in Fig.~\ref{fig:iris}(a). According to the labeled data (open circles), it is obvious that the shortest path that connects all the labels' barycenters is green-red-blue. Therefore, we encode the label by an ordered binary Gray Code as $\{00\}_{\mathrm{Setosa}}$, $\{01\}_{\mathrm{Versicolour}}$ and $\{10, 11\}_{\mathrm{Virginica}} $. We assume that the similarity between arbitrary two data follows a 2-dimensional mixed Gaussian-like function:

\begin{equation}
\begin{aligned}\label{S_iris}
    S_{ij} = \sum_{k=1}^3 \frac{p(k)}{2\left[1-{\left(\rho^{(k)}\right)}^2\right]}
    \left[ \frac{x_{i(1)}-x_{j(1)}} {{\left(\beta_1^{(k)}\right)}^2}+\frac{x_{i(2)}-x_{j(2)}}{{\left(\beta_2^{(k)}\right)}^2}\right. \\
    \left.-\frac{2\rho{(l)}(x_{i(1)}-x_{j(1)})(x_{i(2)}-x_{j(2)})}{\beta^{(k)}_1\beta^{(k)}_2} \right],
\end{aligned}
\end{equation}
where $x_{i(1)}$ ($x_{i(2)}$) is the first (second) element of data $\mathbf{x}_i$, $\rho^{(l)}$ is the correlation coefficient between $\mathbf{x}_{i(1)}$ and $\mathbf{x}_{i(2)}$. In this exmaple, the data have been processed with PCA ~\cite{kessy2018optimal}, so we can assume that $\rho = 0$. $J_{ij}$ are calculated using Eq. \ref{coupling matrix} with $\xi=6$.

In Eq. \ref{S_iris}, $\beta_*^{(k)}$ are the parameters to be learned using the data of the $k$-th label. For each $k$, we substitute $J_{ij}$ into Eq. \ref{L-function} and \ref{P-function} to have:
\begin{equation}\label{L-iris}
\begin{aligned}
    \mathcal{L}\left(\mathcal{D}_l;\beta_*^{(k)}\right) 
    &= -\log p\left(lab_1,\dots,lab_l|\mathbf{x}_1,\dots,\mathbf{x}_l;\beta_*^{(k)}\right)\\
    &= \frac{1}{lZ(\theta)}\sum_a^\alpha\sum_{i,j=1}^l\sum_{k=1}^3\frac{p(k)}{2}\left[\frac{x_{i(1)}-x_{j(1)}}{{\left(\beta_1^{(k)}\right)}^2T}+\right.\\
    & \left.\frac{x_{i(2)}-x_{j(2)}}{{\left(\beta_2^{(k)}\right)}^2T}\right]lab_i^{(a)}lab_j^{(a)},
\end{aligned}
\end{equation}
in which $lab_i^{(a)}$ is the $a$-th bit of $lab_i$ with binary encoding.
Then we can obtain the optimized $\beta_*^{(k)}$ with Eq. ~\ref{P-learning}. 

The classification results are shown in Fig. \ref{fig:iris}(b)-(d): when the 30\% of the data set is unlabeled, the accuracy of the algorithm is 100\%. An accuracy of 94.26\% can still be maintained when 80\% unlabeled data is considered. 

\subsection{Example 2: handwriting digital pictures}
The second example is the handwritten digital recognition using the database from MNIST ~\cite{garris1994nist}. We pick 200 images of digits 1, 8, 5, and 6 from the original data set and reduce the original dimension to 2~\cite{tenenbaum2000global} as shown in Fig. \ref{fig:handwriting digits}(a). According to their barycenters on the manifold, we encode the 4 labels by $\{00\}_1$, $\{01\}_8$, $\{11\}_5$ and $\{10\}_6$.

Here the Euclidean distance given by Eq. \ref{Similarity-Eucli_Dis} and $\xi = 4$ are applied to calculate the similarity matrix $S$ and coupling parameters $J$. Parameters concerning the similarity calculation are trained using similar approaches as the first example.

Fig. \ref{fig:handwriting digits}(b)-(d) shows the classification results. The accuracy of QA-SSL changes from 96.15\% to 92.13\% , as the portion of the unlabeled data in the whole data set increases from 30\% to 80\%, showing the feasibility of this method.

\section{conclusion}

So far, quantum machine learning algorithms have been studied extensively on clustering (unsupervised learning) ~\cite{horn2001algorithm,roitblat2011identifying,kurihara2014quantum,kumar2018quantum} or supervised learning  classification algorithms~\cite{rebentrost2014quantum, li2015experimental}. In this paper we introduce a new semi-supervised learning method based on quantum annealing. In this method, the classification problem is mapped to the QA Hamiltonian through a graph representation, of which the vertices are efficiently implemented by qubits with an encoding scheme based on a binary Grey code. Calculations of the similarity between data are improved with a learning process using various models. Compared with previous proposed classification method using QA, this scheme significantly saves the quantum resources while maintaining the ability to express the original problem. The results of two proof-of-principle examples indicate that this method can still yield high accuracies for classification problem when the amount of labeled data is limited.

\nocite{*}
\bibliography{references}%


\end{document}